\documentclass{elsart}
\usepackage[numbers]{natbib}
\usepackage{graphicx}
\def\aap{A\&A\,  }
\def\acp{Anal. Cell. Pathol. } 
\def\cjaa{Chinese J. Astron. Astrophys.  }
\def\eup{Europhys. Lett.  }

\def\jpc{J. Phys. C  } 
\def\jsp{J. Stat. Phys  } 
\def\jcp{J. Comput. Phys.  } 
 
\def\oe{Optic  Express }
\def\npb{Nuc. Phys. B   }

\def\pla{Phys. Lett. A   }
\def\prb{Phys. Rev. B   }
\def\pre{Phys. Rev. E   }
\def\prl{Phys. Rev. Lett.    }
\def\physa{Phys. A    }
\def\rmp{Rev. Mod. Phys.  }
\def\za{Z. Astrophys.  } 

\begin         {document}
\begin         {frontmatter}
\title 
{
On the statistics of area size
in two-dimensional thick Voronoi Diagrams
}
\author      {Mario Ferraro and Lorenzo Zaninetti \corauthref{cor1}}
\corauth[cor1]{Corresponding author}
\address    {Dipartimento  di Fisica ,
 via P.Giuria 1,\\ I-10125 Turin,Italy }
\ead {zaninetti@ph.unito.it}
\ead [url]{http://www.ph.unito.it/$\tilde{~}$zaninett}

\begin {abstract}
Cells of Voronoi diagrams in two dimensions
are usually considered as having edges of
zero width.
However, this is not the case in several experimental situations
in which the thickness of the edges of the cells is relatively large.
In this paper, the concept of a thick Voronoi tessellation, that is with edges of non-zero width, is introduced and the  statistics of cell
areas, as thickness changes, are analyzed.

\end  {abstract}
\begin {keyword}
02.50.Ey ,  Stochastic processes                        ;
02.50.Ng ,  Distribution theory and Monte Carlo studies ;
89.75.Kd ,  Patterns                                    ;
89.75.Fb ,  Structures and organization in complex systems
\end   {keyword}
\end {frontmatter}

\section{Introduction}

Voronoi tessellations provide a powerful method for subdividing
space in random partitions and have been used in a such diverse fields as
statistical mechanics  \cite{Hentschel2007}, quantum field theory \cite{Drouffe1984},
astrophysics  \cite{Zaninetti2006},
structure of matter  \cite{Jerauld1984_a},  \cite{Dicenzo1989}, \cite{Mulheran2000}, \cite{Pimpinelli2007},
biology \cite{Ryu2007} and medicine \cite{Sudbo2000}.

Algorithms to generate Voronoi diagrams can be
found, among others, in
\cite{Tanemura1983}, \cite{Kumar}, \cite{Okabe2000} 
\cite{Tanemura2003};
usually Voronoi partitions are obtained by considering a
uniform spatial distribution of centers,
so that the probability that there are $n$ points
in a given domain of space obeys a
Poisson distribution,
of constant intensity $\lambda$ \cite{Okabe2000}.
Non Poissonian distribution of seeds have been used in  
\cite{icke1987}, in an astrophysical framework, and examples  of area distribution for this case  
can be found in \cite{Zaninetti2009c}.

Usually, when generating Voronoi diagrams, the thickness of the edges
is not taken into account, and, hence, its influence on the
size of the cells is not considered.
However, in nature, configurations occur that approximate Voronoi tessellations in which
the width of edges is not negligible; examples can be found in the generation of diffraction patterns
\cite{Giavazzi2008}, in the compaction of granular matter
\cite{Slotterback2008} and
crystallization of granular fluids \cite{Reis2006},
and in animal coat formation \cite{Jonathan1981,Koch1994}.

In the next section probability distribution used to fit areas of $2D$ Poissonian Voronoi tessellation will be briefly reviewed; 
next  the statistics of Poissonian thick Voronoi diagrams,
that is with an edge of non-zero width,
will be studied in the two-dimensional case and, in particular, the dependence of
the mean and variance of the area distribution on the thickness of edges will be analyzed.
Furthermore,
a probability density function will be considered, adapted from that proposed in
\cite{Ferenc_2007}, to fit histograms of cell areas.

\section{Probability distributions}

Consider a $2D$ Poisson Voronoi Diagram (PVD for short) and let 
$a_0$ be the area of  its cells.  

No exact solution is known  for the distribution of  
2D Voronoi cells, however there exist several PDF, 
based of the Gamma distribution, that provide an 
approximate solution, with different degrees of 
"goodness": the most general way is via a $3$ 
parameter generalized 
Gamma distribution 
\begin{equation}
G(x;a,b,c) = 
\frac{
a{b}^{{\frac {c}{a}}}{x}^{c-1}{{\rm e}^{-b{x}^{a}}}
}
{
\Gamma  \left( {\frac {c}{a}} \right) 
}
\quad  .
\end{equation}
\cite{Hinde1980}, \cite{Tanemura2003}; 
here $x=a_0/\langle a_0 \rangle $, 
where $\langle a_0 \rangle$ is the area average. 
The previous PDF can be simplified inserting 
$a=1$ and  the following 
 $2$ parameters Gamma PDF  is obtained    
\begin{equation}
g(x;b,c)= 
\frac
{
{b}^{c}{x}^{c-1}{{\rm e}^{-bx}}
}
{
\Gamma  \left( c \right)
}
\label{eq:tanemura}
\quad ,
\end{equation}
\cite{Kumar,Tanemura2005}.

Note that the most recent numerical estimates of the 
parameters $c$ and $b$, in the $2D$ case, 
show that they are very close, 
namely    $b$=3.52418  and  $c$= 3.52440  \cite{Tanemura2005}.
This suggests the use of a one-parameter   Gamma distribution,

\begin{equation}
 p(x;c) = \frac {c^c}{\Gamma (c)}
x^{c-1}
\exp(-cx)
\quad ;
\label{kiang}
\end{equation}

such a distribution has been used by Kiang in his  seminal 
work on Voronoi diagrams \cite{Kiang}.
Note, however, that in \cite{Kiang} it is has 
been assumed  $c=2d$ where $d$ is 
just the dimensionality of the cells, 
so in the present case case $c=4$. 

Finally   
in \cite{Ferenc_2007}, a simpler  distribution has been proposed with no free parameters,
that can be obtained from (\ref{kiang}) by setting
$c=(3d+1)/2$, that is $c=3.5$ in $2D$:

\begin{equation}
f(x;d) =  {C}x^{\frac {3d-1}{2}} \exp{(-(3d+1)x)}
\quad ,
\label{pdfrumeni}
\end{equation}
where
\begin{equation}
C =\frac{\left [\frac{3d+1}{2} \right ]^{(3d+1)/2}}{\Gamma \left(\frac{3d+1}{2} \right)}
\end{equation}

\begin{figure*}
\begin{center}
\includegraphics[width=10cm]{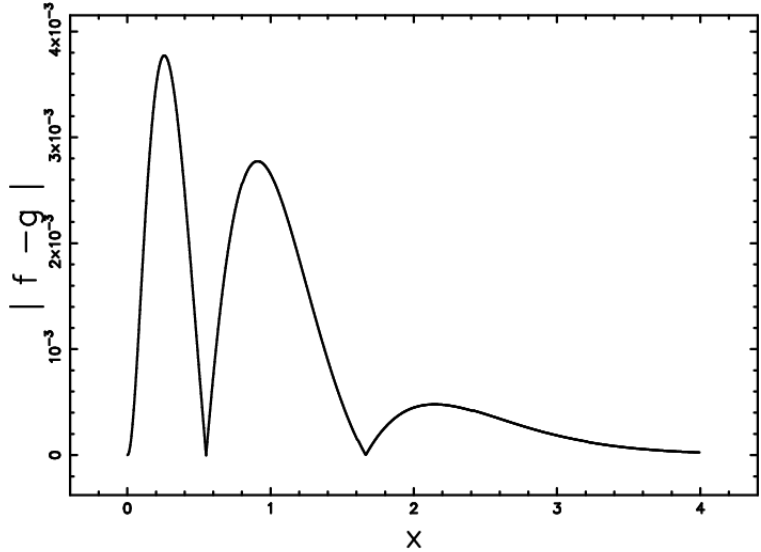}
\end {center}
\caption{
The absolute value of the difference between $f(x;d)$ and 
$g(x;b,c)$
as a function of $x$ when $ d=2$,
$b$=3.52418  and  $c$= 3.52440.
}
\label{f01}
    \end{figure*}

A detailed comparison between $f(x;d)$ and $G(x; a,b,c)$ can be found in 
\cite{Ferenc_2007}.
The usefulness of these distributions in physical problems is determined 
by the trade off between the accuracy with which they fit the data and  their complexity.
We have computed $\delta=|f(x;d)-g(x;b,c)|$ the absolute value of the difference between $f(x;d)$ and 
$g(x;b,c)$, for the $2D$ and the resulting plot is shown 
in Fig. \ref{f01}: the maximum value of $\delta$ is $0.0037$.
Obviously, the PDF $g$ can give a better fit of the data, however $f$ is simpler and  we think that, 
at least  for the purposes of this note,  provides a good enough approximation. 

For future reference we give 
here the variance  and mode of (\ref{pdfrumeni})
which are, respectively,  
\begin{equation}
\sigma^2=\frac{2}{3d+1}
\label{variancerumeni}
\end{equation}

and 
\begin{equation}
\label{mode}
a_m =  \frac { 3d -1} {3d+1}
\quad .
\end{equation}

\section{Two-dimensional thick Voronoi diagrams}
\label{thick}

Let $s$ be the thickness of edges of 2D cells  in a Voronoi tessellation,
denote with $a(s)$ the cells area and with  $a_0$ the area when $s=0$.

The  analysis of area size as a function of $s$ can be made  independent of the area $A$  
of the $2D$ domain $\mathcal A$ in which the Voronoi polygons are generated,
by introducing
a dimensionless parameter
\begin{equation}
\label{ro}
\rho = \frac{s}{\langle a_0 \rangle^{1/2}}=
\frac{s}{\left (\frac{A}{n} \right )^{1/2}}
\quad .
\end{equation}
where  $n$ is the number of seeds of the diagram.

\begin{figure}
\begin{center}
\includegraphics[width=10cm]{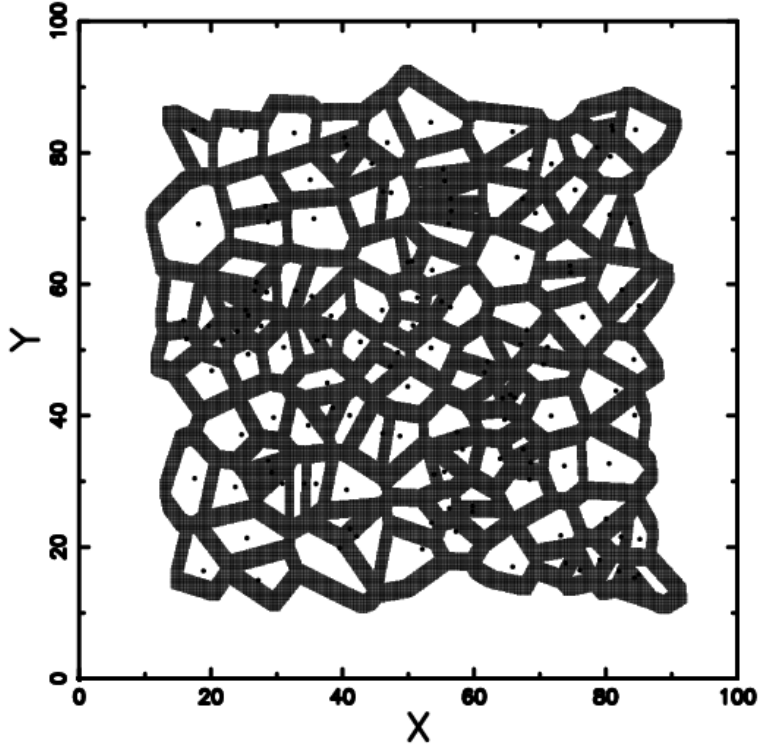}
\end {center}
\caption {
Thick Voronoi diagrams in 2D with $\rho= 0.2$.
The selected region comprises 150  random seeds
marked by a point.Here, for simplicity
$\langle a_0 \rangle =1$.
} \label{f02}
\end{figure}

An example of thick Voronoi diagrams,
with $\rho=0.2$, is presented in
Fig. \ref{f02} where, for illustrative purposes, just $n=150$ centers have been used;
 all simulations presented in the
following have been carried out with $n=4 \cdot 10^4$ centers.

Simulations were run on a LINUX -$2.66$GHz processor: 
Poisson Voronoi tessellation were generated by 
sampling independently the coordinates along  $X$  and $Y$   axes 
from a uniform distribution by means of the subroutine  
RAN2  described in  \cite{press}.
In order to minimize boundary  effects
introduced by cells crossing the  boundary
 of the domain  $\mathcal A$, a square $\mathcal B$ is defined 
larger by a factor $1.5$ and containing $\mathcal A$.
The seeds are placed in the whole $\mathcal B$ domain;   
 furthermore only cells
that do not cross the boundary of $\mathcal A$ are considered,
see  Figure \ref{f02}.

Further information on the code used here can be found in 
\cite{Zaninetti1991}.
The CPU time running time  was $1.42$ s for a seed.

It is obvious that increasing values of $\rho$ make more likely
the occurrence of cells completely covered by edges and indeed, 
from Fig. \ref{f02}, it can be seen that 
some of the smallest cells are completely  covered.
Moreover it is also apparent  for $\rho=0.2$  a relatively large area 
of the cell is occupied
by  edges and this is confirmed by the results presented in  the sequel.
We have then restricted  our analysis to $0 \leq \rho \leq 0.2$.


In general the area $a$ 
 can be considered to be obtained
from $a_0=a(0)$ by the action
of a mapping $K$, so that $a(\rho)=K(a_0, \rho)$ rescales the cell size.
Then $K$ can be seen as a nonlinear scaling
operator and can be given in a form akin to that used for 
the group of scaling \cite{blumankumei},
namely
\begin{equation}
a(\rho) =k(a_0, \rho)a_0.
\label{pass1}
\end{equation}

Note that $k$ is a decreasing function of $a_0$,  in that large cells are relatively less affected
than small ones by
the occurrence of an edge of width $s$.

Since cells are  irregular polygons it is difficult to compute an explicit form of $a$ and $k$ 
since it depends on the shape of the cell,  however
some general properties are readily apparent, which suffice for our purposes.
The area $a$ can not contain powers of $s$, and hence of $\rho$, larger than $2$, for reasons of dimensional consistency, and that holds for $k$ too  
(compare Eqs. (\ref{pass1}));  
it is  also obvious that

\begin{equation}
k(a_0, 0 )=1, \qquad dk(a_0, \rho)/d\rho|_{\rho=0} < 0.
\label{relations}
\end{equation}

Then $k$ can be written as
\begin{equation}
k(a_0, \rho)=1-h(a_0)\rho + \frac{1}{2}g(a_0)\rho^2,
\label{kappa}
\end{equation}
where
$$h(a_0) =\mid dk(a_0, \rho)/d\rho\mid_{\rho=0} \mid, \qquad  
g(a_0)=  d^2k(a_0, \rho)/d\rho^2\mid_{\rho=0},$$

and the area $a(\rho)$ is 

\begin{equation}
a(\rho)=a_0 \left  ( 1-h(a_0)\rho + \frac{1}{2}g(a_0)\rho^2 \right ),
\label{arearho}
\end{equation}

with the understanding that $a(\rho)$ is set to $0$ if Eq.(\ref{arearho}) yields a value less then $0$.

Since $\rho$ is small, only the linear term in (\ref{kappa}) needs to be considered,
then 

\begin{equation}
a(\rho) =a_0 \left ( 1-h(a_0)\rho+\frac{1}{2}g(a_0)\rho^2\right )\approx a_0-a_0h(a_0)\rho,
\label{pass2}
\end{equation}

from which

\begin{equation}
\langle  a(\rho) \rangle \approx  \langle a_0 \rangle-
\langle a_0h(a_0) \rangle \rho.
\label{pass4}
\end{equation}

\begin{figure}
\begin{center}
\includegraphics[width=10cm]{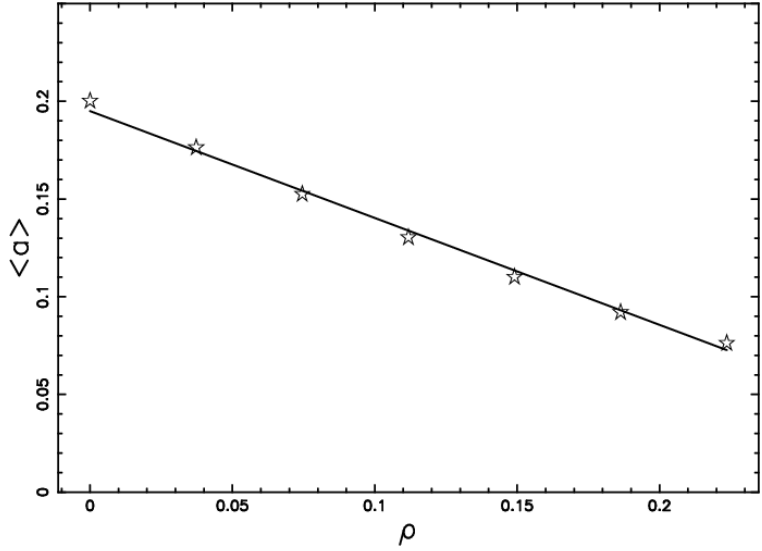}
\end {center}
\caption {
Area average versus thickness.
The stars represent the results of the simulations and
the full line  is  given by Eq. (\ref{pass4}).
} \label{f03}
\end{figure}

A linear decrease of the average area is also
the outcome of simulations,
as shown in Fig. \ref{f03}
where a comparison with a linear fit is presented: values of $\langle a_0 \rangle$ and
$\langle a_0h(a_0) \rangle$ have
been computed by standard fitting procedures
(least squares method) and are reported in
Table~\ref{linear}.
\begin{table}
 \caption[]{Coefficients  of the linear fit. }
 \label{linear}
 \[
 \begin{array}{llll}
 \hline
 \langle a_0 \rangle &
 \langle h(a_0) a_0 \rangle
 & \sigma^2_{a_0} & C \\
 \noalign{\smallskip}
 \hline
 \noalign{\smallskip}
 0.19 &  0.55 & 0.01  & 0.016     \\
 \hline
 \hline
 \end{array}
 \]
 \end {table}

\begin{figure}
\begin{center}
\includegraphics[width=10cm]{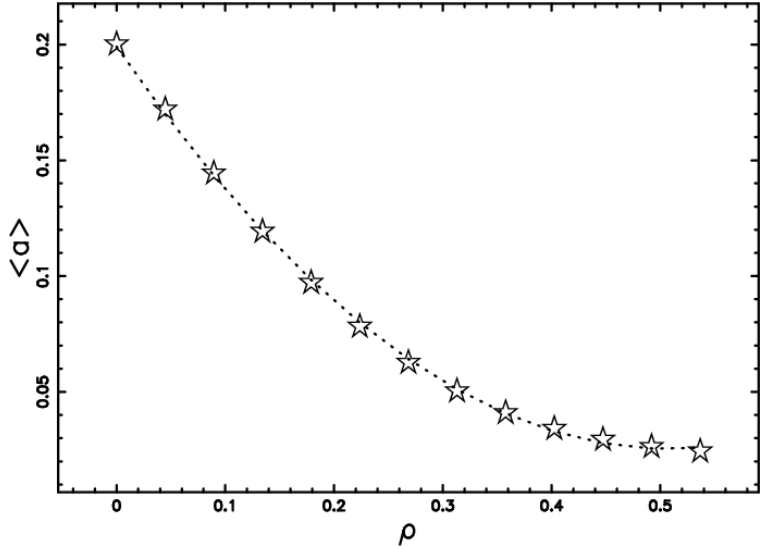}
\end {center}
\caption {
Area average versus thickness.
The stars represent the results of the simulations and
the full line  Eq. (\ref{qfit})  with   
$\langle a=0.2 \rangle$ ,  $\langle h(a_0) a_0 \rangle=0.68$ and ${1}/{2}\langle a_0 g(a_0) \rangle=0.67$.
} \label{f04}
\end{figure}

It should be observed that the area decreases quite sharply even for values of $\rho$ which 
allow the linear approximation of Eq. (\ref{pass2}); 
in particular for $\rho=0.2$, 
the average area is
$\langle a(\rho) \rangle \approx 0.5 \langle a_0 \rangle$.
The trend is linear up to  $\rho =0.25$ approximately, for this value
$\langle a(\rho) \rangle \approx 0.3\langle a_0 \rangle$.  
For illustrative purposes Fig. \ref{f04} shows the trend of $\langle a (\rho) 
\rangle$
in the interval $0  \leq \rho \leq 0.5$, 
 fitted with the equation obtained by averaging the terms of 
Eq. (\ref{arearho}), that is 
\begin{equation}
\label{qfit}
\langle a(\rho) \rangle =\langle a_0 \rangle-
\langle a_0h(a_0) \rangle \rho+\frac{1}{2}
\langle a_0 g(a_0) \rangle \rho^2 \quad .
\end{equation}

The variance $\sigma^2_a$  can then  be calculated:
\begin{equation}
\sigma^2_a=\left < ( a-\left < a \right >)^2 \right > =
\left < [(a_0-a_0h(a_0)\rho)-\langle( a_0 - a_0h(a_0)) \rangle \rho ]^2\right >,
\label{passv1}
\end{equation}

from which it is straightforward to obtain:
\begin{eqnarray}
\sigma^2_a &=& \left <( a_0-\langle a_0 \rangle )^2 \right >
-2\left < (h(a_0)a_0-\langle h(a_0)a_0 \rangle )(a_0-\langle a_0 \rangle) \right > \rho \nonumber \\
&+&
\left < (h(a_0)a_0-\langle h(a_0)a_0 \rangle )^2 \right > \rho^2,
\label{passv2}
\end{eqnarray}

and, by making use again of the linear approximation,

\begin{equation}
\sigma^2_a \approx \left <( a_0-\langle a_0 \rangle )^2 \right >
-2\rho \left < (h(a_0)a_0-\langle h(a_0)a_0 \rangle )(a_0-\langle a_0 \rangle) \right > .
\label{passv3}
\end{equation}

Set
$$ \left < (h(a_0)a_0-\langle h(a_0)a_0 \rangle )(a_0-\langle a_0 \rangle) \right > =
C(h(a_0)a_0, a_0),$$

then Eq.(\ref{passv3}) can be written
as

\begin{equation}
\label{passvqf}
\sigma^2_a \approx \sigma^2_{a_0}-2C(h(a_0)a_0, a_0)\rho.
\end{equation}

\begin{figure}
\begin{center}
\includegraphics[width=10cm]{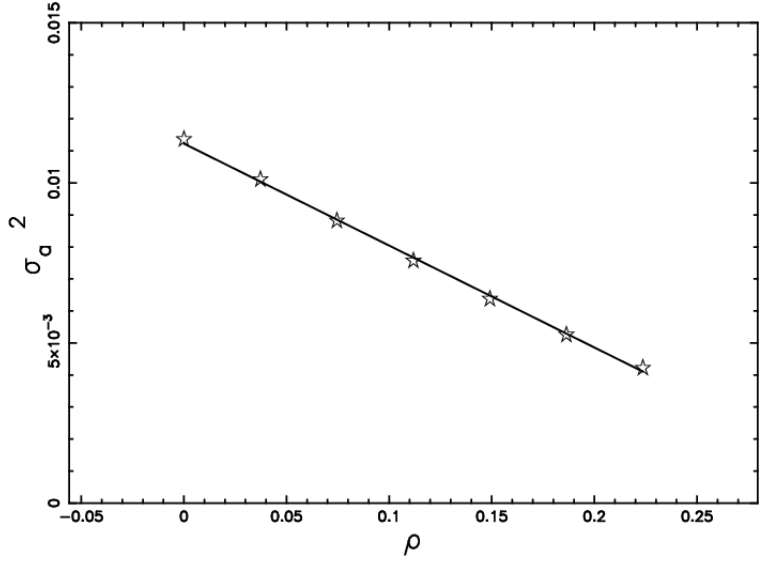}
\end {center}
\caption {
Variance of cells area versus thickness.
The stars represent the results of the simulations and
the  line the theoretical variance,
as given by formula~(\ref{passvqf}).
} \label{f05}
\end{figure}

Now $C(h(a_0)a_0, a_0) > 0$, because it is the covariance of $h(a_0)a_0$ with $a_0$
and, by definition, both $a_0$ and  $h(a_0)$ are positive; therefore, if the approximation of
Eq. (\ref{passv3}) holds,
$\sigma^2_a$ must fall off linearly.
Comparison between  simulations
and the results of
Eq. (\ref{passvqf}) are shown in Fig. \ref{f05} and it is clear
from the figure that
the  effect of a quadratic term on the fit is negligible; indeed we have verified
that $\left < (h(a_0)a_0-\langle h(a_0)a_0 \rangle )^2 \right > $
 is about an order of magnitude smaller than $C(h(a_0)a_0, a_0)$.
Numerical values of
$\sigma^2_{a_0}$ and  $C(h(a_0)a_0, a_0)$ are  shown
in Table~\ref{linear}.

\section{Fitting area distributions of thick Voronoi diagrams} 

We consider now  the probability density function (PDF) of $a$; as noted earlier
PDFs of Voronoi cell size  are commonly expressed in
terms of a standardized variable $x$, which, in the  present case, takes the form  $x=a/<a>$, so that, obviously,
 $\langle x \rangle =1$.
For future reference, we compute the variance
 $\sigma^2_x$ of $x$, which is given by

\begin{equation}
 \sigma^2_x=\frac{\sigma^2_a}{\langle a \rangle ^2},
\label{sigx}
\end{equation}
from which, making use of Eqs. (\ref{pass4}) and (\ref{passvqf}),
\begin{equation}
 \sigma^2_x=\frac{\sigma^2_a}{\langle a \rangle ^2} \approx
\frac{\sigma^2_{a_0}-2C(h(a_0)a_0, a_0)\rho}
{\left < a_0- a_0h(a_0) \rho \right >^2}.
\label{sigmax}
\end{equation}

\begin{figure}
\begin{center}
\includegraphics[width=10cm]{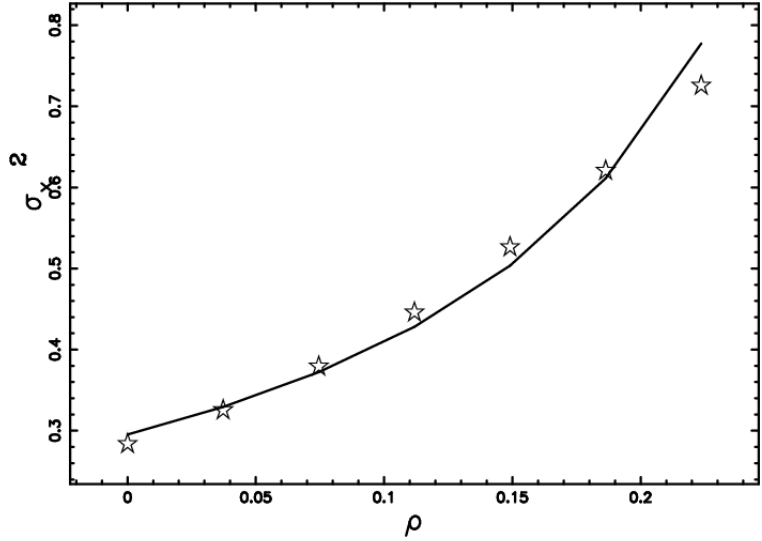}
\end {center}
\caption {
Variance of the  standardized variable $x$  versus thickness.
The stars represent the results of the simulations and
the full line reports the theoretical variance
as given by formula~(\ref{sigmax}).
} \label{f06}
\end{figure}

Values of  $\sigma^2_x$  provided by  Eq. (\ref{sigmax}) are in good agreement with the
results of simulations as shown in Fig. \ref{f06}.

To investigate the distribution of cell areas, we have used the
probability density function 
(\ref{pdfrumeni})  :
however, since $\sigma^2_x$ increases with $\rho$,
from Eq. (\ref{variancerumeni}) it is clear that, in order to use the PDF $f$ to fit
histograms of simulated data, $d$ must be considered to be a variable parameter which decreases for
increasing $\rho$. From Eqs. (\ref{variancerumeni}) and (\ref{sigmax}),
it is straightforward to derive a formula for $d$:
\begin{equation}
\label{trendd}
d(\rho)=\frac{1}{3}\left (\frac{2}{\sigma_x^2}-1\right )
=\frac{1}{3}\left [\frac{2 \left < a_0- a_0h(a_0) \rho \right >^2  }
{\sigma^2_{a_0}-2C(h(a_0)a_0, a_0)\rho} -1 \right ].
\end{equation}

\begin{figure}
\begin{center}
\includegraphics[width=10cm]{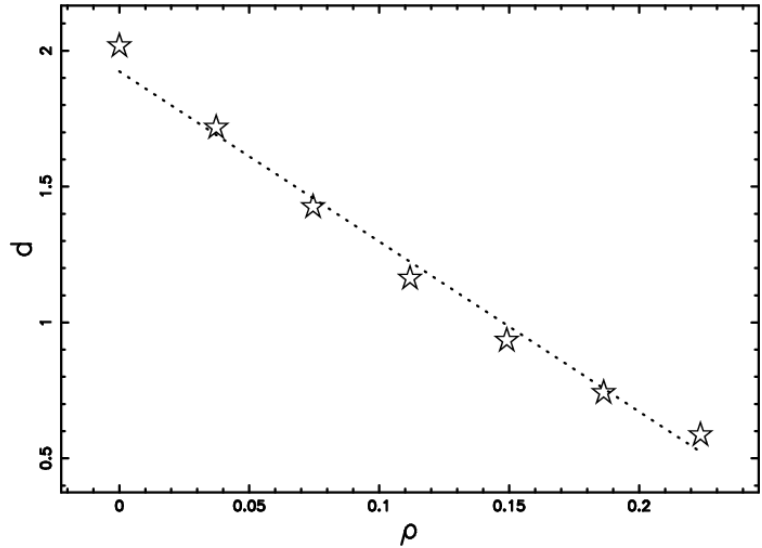}
\end {center}
\caption
{
The dimension $d$ that models the Voronoi  cell
standardized area-distribution in 2D as function
of the adimensional parameter $\rho$  (dotted line),
see  formula~(\ref{trendd})
and simulated points (stars) .
}
\label{f07}
\end{figure}

\begin{figure}
\begin{center}
\includegraphics[width=10cm]{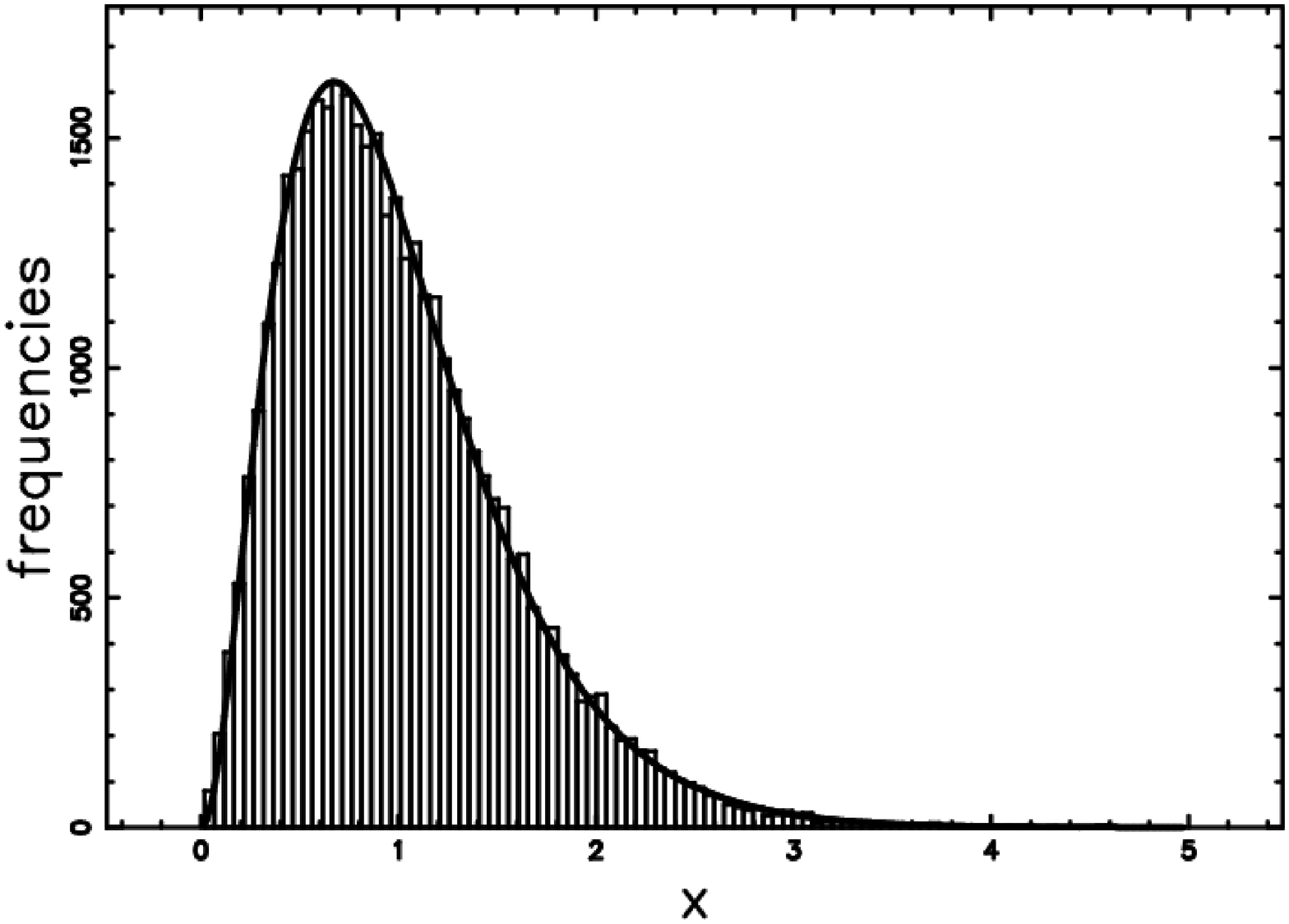}
\end {center}
\caption
{
Histogram (step-diagram)  of
the Voronoi  normalized   thick area distribution
in 2D
with a superposition of the
gamma PDF derived by Eq. (\ref{pdfrumeni}), with 
$d$ given by Eq. (\ref{trendd}).
The number of seeds and bins are  40000  and 
100 respectively: here 
$d   =1.71 $                  ,
$\rho=0.04$                   ,
$\chi^2=151 $
$\chi_{\nu}^2=1.53 $              .
}
\label{f08}
    \end{figure}

Empirical values of $d$ have been found
by the method
of matching moments and are shown in Fig. \ref{f07}, together with the fit
provided by Eq. (\ref{trendd}).
We have generated histograms of area distribution
for different values of $\rho$
and have used the PDF $f$ with the corresponding parameter $d$ derived with
Eq. (\ref{trendd}): statistical tests show that  $\chi^2_{\nu}$ increases 
with $\rho$ and that
$\chi^2_{\nu} \leq 1.53$  up to  $\rho=0.04$
(see Fig. \ref{f08}),
thus the fit 
is   adequate only  for
very  small  values of $\rho$.

Even though the PDF $f$, with $d$ given by Eq. (\ref{trendd}),
 gives good results only for small
$\rho$ values, it can be used to
predict, at least qualitatively, the change in shape of
the empirical distribution.
The decrease of the parameter
$d$  implies a shift
of the mode $a_m$ toward zero (see Eq.( \ref{mode})) and
that occurs also in the histograms generated by
the simulations.

\begin{figure}
\begin{center}
\includegraphics[width=10cm]{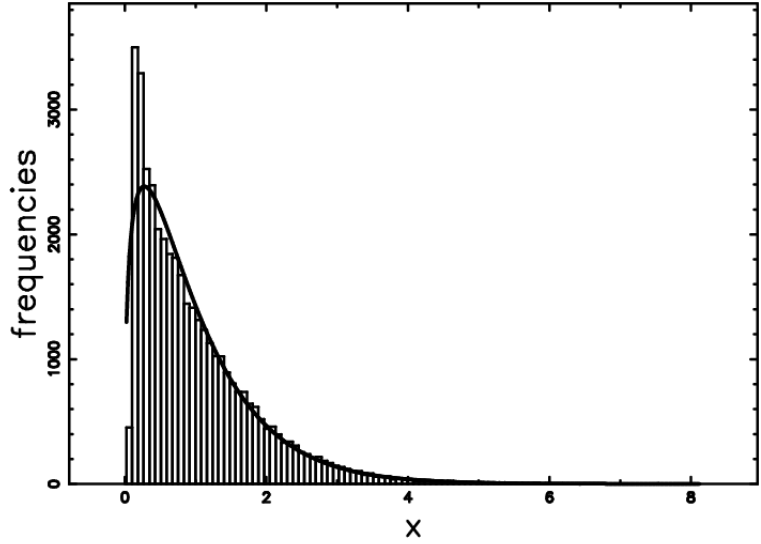}
\end {center}
\caption
{
Histogram (step-diagram)  of
the Voronoi  normalized   thick area distribution
in 2D
with a superposition of the
gamma PDF  as derived  by equation~(\ref{pdfrumeni}) with 
$d$ given by Eq. (\ref{trendd}).
The number of seeds  and bins are 40000 ,
and 100, respectively, and
$\rho=0.22$,
$\chi^2=5049 $
$\chi_{\nu}^2=51 $              .
}
\label{f09}
    \end{figure}

The result is the appearance
of a PDF  decreasing monotonically with $a$,
as an example see Figure~\ref{f09}, where is clear that lack of agreement with 
the modified PDF $f$.

\section{Conclusions}

Voronoi Diagrams in 2D are usually generated
as irregular polygons whose edges, in principle,
have zero thickness: however
in several experimental situations, Voronoi cells appear to have
edges of relatively large width.

Clearly the emergence of thick edges
is related 
to the formation of
configurations representing approximate 
Voronoi diagrams, as results  
of chemical and phy sical mechanisms, are 
in general quite complex.

Consider, for instance, pattern formation in 
certain animals coats (e.g. giraffe) by 
reaction diffusion processes.
Here two-dimensional Voronoi diagrams are generated by an 
activator $a$, diffusing 
from  randomly placed point sources, which  switches on 
the production of melanin,  
this switch being controlled by a  threshold $\theta$
\cite{Jonathan1981}; 
in  a  more complex model  \cite{Koch1994} 
melanin production is modulated by 
the concentration of a substrate $s$. 
Thickness of edges is then determined by the value of $\theta$ 
\cite{Jonathan1981}, or by the abundance and removal rate of 
the  substrate \cite{Koch1994}.

Voronoi cells  can also be generated  
when an homogeneous medium is occupied 
by domains 
emerging from  random placed seeds with the same isotropic growth rate \cite{Pineda2007,Pineda2008}: 
such is the case of crystals   \cite{Pineda2007,Pineda2008} or bubbles in volcanic eruptions 
\cite{Blower2002}. In this case the width of the edges can be determined by the relation between 
the amount of growth and the size of the domain where it takes place. 

Edges  width should  be constant, 
at least approximately, when  processes leading to 
the formation of Voronoi cells are symmetric, 
whereas if symmetry breaks down edges of different thickness 
must be expected. Consider again animal coat formation:
if the  threshold $\theta$ is not constant 
over the domain where cells are formed 
edges of different width will emerge. The same effect can result if 
the constant value of $\theta$ is replaced by a probability distribution $p(\theta)$.

Here a  general method has been presented to compute the statistics of cell areas as
the thickness parameter $\rho$ varies: here, for  each value of $\rho$, edges have the same width.  
Theoretical computations as well as results of simulations show that the
mean area $\langle a \rangle$ and variance $\sigma^2_a$
fall off linearly for  $\rho$ increasing in the interval $[0, 0.2]$:
in particular the mean area shows   a marked decrease and  at $\rho=0.2$ 
is reduced to $50 \%$ of its original value.

We have tried to fit the  simulated distribution of the standardized
variable $x=a/\langle a \rangle$ 
with the PDF presented in \cite{Ferenc_2007}, by using $d$ as  a free parameter, 
but such a fit holds only for   $\rho \leq 0.07$; for larger values of 
$\rho$ simulations show that the mode shifts close to zero more rapidly than predicted by 
equation Eq. (\ref{mode}).

Finally, it should be noted that the formation of thick edges  can be seen as
a particular example of a process by which cells are eroded.
The approach presented here, however, is general enough to be readily adapted
to analyze the
statistics of cell areas for different cases of cell erosion, for instance as in diffusion-limited aggregation of Voronoi diagrams \cite{Mulheran2000},  in that every erosion operator
$k$ must have the form given by Eq. (\ref{kappa}) and the terms $h$ and $g$ can be
determined from the data, experimental or simulated.

\section*{Acknowledgments}
We thank the referees   for  valuable criticism and suggestions.


\end {document}